\documentstyle[12pt,epsfig]{article}

\def\proof{\par\vskip 3pt\noindent\hbox{\bf Proof:}\quad}
\newtheorem{thm}{Theorem}
\newtheorem{lem}{Lemma}
\def\qed{\relax\ifmmode\hskip2em \Box
\else\unskip\nobreak\hskip1em $\Box$\fi}
\def\Pscr{{\cal P}}
\def\Rset{\hbox{I\hskip-0.23em R}}

\begin{document}

\title{Optimal Cooperation and Submodularity for Computing
Potts' Partition Functions with a Large Number of States}

\author{J-Ch. Angl\`es d'Auriac\footnote{
CNRS-CRTBT B. P. 166, F-38042 Grenoble, France}\\
F. Igl\'oi\footnote{
Research Institute for Solid State Physics and Optics
H-1525 Budapest, P.O.Box 49, Hungary
and Institute for Theoretical Physics,
Szeged University, H-6720 Szeged, Hungary}\\
M.Preissmann and A.Seb\H{o}\footnote{Laboratoire LEIBNIZ-IMAG
46, Avenue F\'elix Viallet, 38000 Grenoble Cedex, France}
}

\maketitle

\begin{abstract}
The partition function of the $q$-state Potts model with random
ferromagnetic couplings in the large-$q$ limit is generally
dominated by the contribution of a single diagram of the
high temperature expansion. Computing this dominant diagram
amounts to minimizing a particular submodular function.
We provide a combinatorial optimization algorithm,
the optimal cooperation algorithm,
which works in polynomial time for any lattice.
Practical implementation and the speed of the method is also discussed.
\end{abstract}

\section*{Introduction}
Spin models with discrete symmetry are often used to describe
order-disorder transitions, such as in magnetic systems or in
absorbed monolayers, etc, for a review see in\cite{yeomans}.
In two-dimensional regular lattices
and with homogeneous couplings there are some models (Ising
model, critical Potts model, eight-vertex model, etc.) with an
asymptotically exact solution for the partition function in the
thermodynamic limit\cite{baxter}. In the presence of quenched, that is time
independent, disorder this type of exact solutions are scarce
which has greatly hampered our understanding about collective
phenomena in disordered systems, for example about the properties of
spin glasses.

For disordered systems, such as for spin models with random
couplings and/or fields one usually performs an exact computation
on finite lattices and the results are then extrapolated.
An important limitation of this finite lattice method is the
available sizes, which can be treated by different computational
methods. For most of the discrete spin models a computation
of the ground state represents a hard optimization problem
when there is frustration and
it is even much more complex to compute the partition function.

In this respect the Ising model, which is equivalent to the Potts
model with $q=2$ states, is an exceptional problem since
the ground (minimum energy)
states can be found in polynomial time using the solution of max
cut problems in planar graphs that involves matching theory
\cite{Bspin} \cite{9} \cite{MRAdA}.
The full partition function of the Ising problem
can also be computed for lattices of bounded genus \cite{GLV}.
Another example of application of optimization theory
to lattice statistical mechanics is the Random Field Ising Model
where the ground state can be found in polynomial time
in any dimension \cite{JPhysLett85} \cite{EuroPhys97}.
A discussion about optimal cuts in statistical mechanics
can be found in \cite{JouCompMod97} and a
general review about the application of combinatorial optimization
methods in statistical physics can be found in\cite{admr}.

For the $q=3$-state Potts model, when the number of states is three,
the computation of the minimum
energy of the system involves (NP-)hard combinatorial problems
such as minimizing the 3-cut in a graph, and no efficient
procedure is known for computing the partition function.

If the number of states can be arbitrary large (and given as part of the input)
the partition function has an exponential number of terms, and
the problem may look hopeless.
In this paper we show that this is not the case: when
the number of states tends to infinity,
the partition function can be determined up to
an integer multiplicative factor, and some average
properties of the system can be inferred. This evaluation
goes trough the minimization of a submodular function.
Note that in \cite{FIS} a simulated annealing method and an
approximate combinatorial method have been used. The
solution we give here is both exact and much more efficient.  

Even though there are ready-made algorithms for general
submodular function minimization \cite{GLS2} \cite{S}
\cite{FFI} \cite{I}, the resources they require are
far beyond the possibilities of the present application. While
the existing first implementations of general submodular function
minimization may require too much time already to solve the
optimization problem on a lattice containing a dozen
of sites (vertices of the input graph
corresponding to the variables of the general problem),
we want and will solve problems involving
more than two hundred thousand `sites'.
For this we use methods inspired  by \cite{FT} \cite{CuO}
\cite{B} \cite{BBM}
  that fit the
particularity of the application and develop an algorithm which
is tailored to graphic submodular functions. Besides the
computational advantages of this approach, we are also aiming at
conceptual simplicity and elegance. Readers who seek for more
mathematical background and connections can consulte \cite{AIPS}.

In Section~\ref{sec:ph} we describe the  problem of physics and
show how it reduces to a function maximization.
In Section~\ref{sec:pr} we formulate the mathematical problem, the
preliminaries  and the broader context of our algorithm; in
Section~\ref{sec:ext} and ~\ref{sec:PQPW} we show how to construct an
optimal solution and  in
Section~\ref{sec:alg} we state the algorithm.
Finally, in Section~\ref{sec:res} we analyze the practical implementation
of the method and display a few computational results.
  The paper is self-contained, the algorithms and their correctness will be
fully proved.

\section{The problem in terms of physics.}
\label{sec:ph}
For a review of the huge amount of work devoted to the Potts
model see \cite{FWu}. In this section we recall the Potts model
and show why its partition function (up to an integer multiplicative factor)
is determined, if the number of `states' tends to infinity
by the minimum value of a submodular function.

A {\em lattice} of spins is given at each {\em site} of which
is a variable. We denote the number of sites by $n$. Each
variable $\sigma_1,\ldots,\sigma_n$ can take
values in \(\{0, 1\ldots, q-1\}\) for a given integer $q$.
Pairs of neighboring  sites on the lattice are called {\em
bonds}, the number of
bonds by $m$.

Each configuration $\sigma=(\sigma_1,\ldots,\sigma_n)$
has an energy
\begin{equation} \label{defE} E(\sigma)=\sum _{ij}K_{ij}\delta
_{\sigma _{i}\sigma_{j}},
\end{equation}
where \( \sigma _{i}\in \{0,\ldots, q-1\}\)
the sum runs over all bonds $ij$, \( K_{ij}\in\Rset_+\)
is a given non-negative weight of bond $ij$, and \(\delta_{ab} \)
is the Kronecker symbol ($1$ if $a=b$ and $0$ otherwise). For
mathematicians: a lattice is a graph, sites are vertices, bonds
are edges.  The aim is to compute or approximate the partition
function
\begin{equation} \label{partF} Z(K_{ij}:
\hbox{$ij$ is a bond})=\sum_\sigma \exp
(E(\sigma))=\sum_\sigma\prod_{ij}e^{K_{ij}\delta _{\sigma
_{i}\sigma _{j}}},
\end{equation}
where the summation runs over all assignments of values
$\sigma=(\sigma_1,\ldots,\sigma_n)$, and the
product over all bonds $ij$ (the inverse temperature
has been included in the couplings $K_{ij}$).

We follow \cite{FIS}.
Note that \( e^{K\delta}=1+(e^{K}-1)\delta \) for \( \delta
\in\{0,1 \} \) ; introduce \(\nu _{ij}=e^{K_{ij}}-1\) and expand
the product of sums: $$\prod_{ij}e^{K_{ij}\delta _{\sigma
_{i}\sigma _{j}}} = \prod_{ij}(1+(e^{K_{ij}}-1)\delta _{\sigma
_{i}\sigma _{j}}) =\sum_F \prod_{ij\in F} \nu_{ij}\delta _{\sigma
_{i}\sigma _{j}},$$ where the summation runs over all subsets of
bonds $F$. Substituting this to (\ref{partF}) we get:
$$Z(K_{ij})= \sum_\sigma\sum _{F}\prod
_{ij\in F}\nu _{ij}\delta _{\sigma _{i}\sigma _{j}}$$
\begin{equation} \label{kasfor}=\sum_F\sum _\sigma\prod _{ij\in F}\nu
_{ij}\delta _{\sigma_{i}\sigma _{j}}=
    \sum _{F} q^{c(F)}\prod _{ij\in F}\nu _{ij},
\end{equation}
where the sum runs over all subsets \( F \) of bonds,
and where \( c(F) \) is the number of connected
components of  \( F \) on the set of all sites, counting also the
isolated sites among the components. (By convention if $F$ is
empty then $\prod _{ij\in F}\nu _{ij} = 1$.) We got the last
equality by counting the number of different
$\sigma=(\sigma_1,\ldots,\sigma_n)$ for which the product is
nonzero; since it is nonzero if and only if $\sigma(i)=\sigma(j)$
for every  $ij\in F$, that is, if and only if $\sigma$ is constant
on every connected component of $F$, all possible $\sigma$ can be
enumerated by choosing an element of $Z_q$ for every connected
component of $F$ independently. Therefore  we have exactly
$q^{c(F)}$ such $\sigma$.

Note that this sum has \( 2^{m} \) terms, and that in
(\ref{kasfor}) \( q \) does not need to have an integer value.
(This provides a way of defining  Potts' model for non-integer
values \cite{KF}.) Clearly, $K_{ij}\ge 0$ is equivalent to $\nu_{ij}\ge 0$,
and we can introduce a new set of variables \(
\alpha _{ij} \) with
\[
\nu _{ij}=q^{\alpha _{ij}},\] and the partition function
becomes\begin{equation} \label{part2} Z=\sum _{F}q^{c(F)+\sum
_{ij\in F}\alpha _{ij}}.
\end{equation}
Finally one introduces the function
\begin{equation}
\label{f} f(F)=c(F)+\sum _{ij\in F}\alpha _{ij}
\end{equation}
so that\begin{equation}
\label{part3}
Z=\sum _{F}q^{f(F)}.
\end{equation}

As pointed out in ref \cite{FIS}, while \( q \) tends to infinity
the sum in (\ref{part3}) is asymptotically equal to  $Z = N q^{ f^\star}$
where $f^{\star}=max_{F \subseteq\{1, \ldots, n\}} f(F)$ and $N$ is
the number of optimal sets.
Note that if the weights $w_{ij}$ are arbitrary reals, the
degeneracy $N$ is likely to be $1$.

We will analyze the
combinatorial optimization problem of finding this maximum.
The problem of finding the degeneracy of the optimal set
will not be addressed. In the analysis of the Potts model
magnetic properties are related to geometrical properties
of the optimal sets $A$, and specifically to
their fractal dimension (if any).
We assume that, on average and for large enough lattice,
this fractal dimension is the same for all sets, and therefore
finding a single optimal set is sufficient for our purpose.
We will present  methods which apply to any set of weights \(
\{\alpha _{ij}\}\) in an arbitrary lattice
(graph) and provide a simply stated and efficient algorithm.

\section{The problem in terms of graphs.}
\label{sec:pr}

For basic graph theoretic notions, terminology and notation we
refer to \cite{Lexercise}.
Given a graph $G= (V, E)$ on $n$ vertices with weight function
$w: E(G)\longrightarrow\Rset$,
we want to solve  the following problem:
$$
\hbox{Maximize}\, \{f_{G,w}(A)=c_{G}(A)+ \sum_{e \in A}w(e):
A\subseteq E(G)\}
\eqno{(POTTS)},
$$
where $c_{G}(A)$ is the number of connected components
of the graph $G(A)= (V,A)$. Note a slight change of notation with
Eq.~\ref{f} where an edge is denoted by $ij$ the two corresponding
adjacent vertices.
We can suppose $0<w(e)<1$ for all $e\in E(G)$, because if $w(e) =
0$, then deleting the edge, if $w(e)\ge 1$, then contracting it
(identifying its two endpoints) we get an equivalent problem.

When no confusion is possible we will simply use $f$ or $f_{G}$ for
$f_{G,w}$, $c$ for $c_{G}$.
This function that has to be maximized in order to solve (POTTS)
has a crucial and well-known property (see for instance
\cite{Lexercise}, Exercise 6.2). For any two subsets $A$ and $B$ of
edges of $G$ : $$ f(A)+f(B) \le f(A \cup B)
+ f(A \cap B).\eqno{(SUPER)}$$
Indeed, it is clear that $\sum_{e \in A}w(e) +
\sum_{e \in B}w(e) = \sum_{e \in A \cup B}w(e) + \sum_{e \in A
\cap B}w(e)$, so we only have to show
that $c(A)+c(B) \le c(A \cup B) + c(A \cap B)$; that is $c$
satisfies the (SUPER) property. For the sake of completeness we
include a proof.
Proceed by induction on $|A \setminus B|+|B \setminus A|$.
If this number is 0 then the statment is obvious. If not, then there
exists say $e \in A \setminus B$. By induction
$c(A \setminus e)+c(B) \le c(A \cup B \setminus e)+ c(A \cap B)$. Deleting
an edge increases the number of connected components by at most 1 and
so we are done unless the equality holds and $c(A \setminus e) = c(A)$
and $c(A \cup B \setminus e) = c(A \cup B) +1$. But this is impossible
since $c(A \setminus e) = c(A)$ means that the two endpoints of $e$ are
in the same connected component of $G(A \setminus e)$, and then also in
the same component of $G(A \cup B \setminus e)$.

A function $g$ defined on the set of subsets of a set $S$, with values in
$\Rset$ is said to be {\it submodular} if $g(A)+g(B) \ge g(A \cup B)
+ g(A \cap B)$ for all subsets $A$ and $B$ of $S$. The (SUPER) property
shows that the function $-f$, whose minimization solves
(POTTS), is submodular.

By now several polynomial algorithms \cite{GLS2} \cite{S}
\cite{FFI} \cite{I} are known to	minimize a submodular function,
therefore we can already claim that there exists a polynomial
algorithm solving our problem (POTTS). However, these general
algorithms are not efficient enough for our instances. Fortunately, 
the specific
properties of our problem make possible the use of a considerably
simpler	and quicker algorithm.
Several more specialized algorithms were already known
(see \cite {AIPS} for an extended presentation) and in
particular Cunningham \cite {CuO} worked out a specialized algorithm
to the	``optimal attack problem" :``The weight of an edge
represents the effort required by an attacker to destroy the
edge, and the attacker derives a benefit for each new
component created by destroying edges	$\ldots$". (POTTS) is
equivalent to the ``optimal attack problem"
but with a `complementary' viewpoint :
{\em the weight of an edge represents the benefits of cooperation
between two vertices (say, researchers), furthermore  there is a unit
support	for each component (say, for each research project).}
So there is a loss of support when two components unite.
Cooperate optimally!

The algorithm described	below has its roots in  \cite{CuO}
and is similar to \cite{B}, \cite{BBM}. It has the best asymptotic
worst case complexity, the same	as \cite{BBM}, and is probably
the simplest, both conceptually	and in the use of computational
resources of the implementation. Sophisticated ingredients
were necessary for finding this solution, but we don't need
to make	explicit use of	them. The interested reader may find more
details on that subject in \cite {AIPS}.

\bigskip
We return now to the (POTTS) problem. An important
observation is the following: if $A$ is a set of edges, and $X$ is
the vertex set
of a connected component of $G(A)$, then adding to $A$ edges
{\em induced} by $X$, that is with both endpoints in $X$,
increases the value of $f$.  Thus in an optimal solution $A^{*}$
of (POTTS) each connected component of $A^{*}$ contains all the
edges of $G$ it induces.
Let us call {\it Potts partition} any
partition of the vertices such that each class induces a connected
subgraph of $G$.

All we have to determine is :
{\it
a Potts partition maximizing the total weight of edges induced by the
classes plus the number of classes.}

As a consequence, following \cite{BBM}, during all the procedure
it will be sufficient to consider subsets of vertices instead of
subsets of edges.

A set $F$ of edges optimizing (POTTS) will be called {\em optimal}, its
associated Potts partition ${\cal P}_{G}(F)= \{X_{1}, \ldots X_{k}\}$
will be called {\em optimal} as well. For any subset $F$ of edges
let $w(F)= \sum_{e \in F}w(e)$ and for any subset $X$ of vertices let
$w(X)= \sum_{e  \in E(X)} w(e)$ where $E(X)$ is the set of edges with
both extremities in $X$.

\section{Extension of a solution}
\label{sec:ext}
In this section we will see how any set of edges maximizing (POTTS)
for a graph $G$
minus a vertex,
may be completed so that it maximizes (POTTS) for the graph $G$ itself.

Let $G$ be a graph with a weight function $w$ on the edges, let $x$ be
a vertex of $G$
and let $G'$ be the subgraph of $G$ obtained by deleting $x$.
The edges of $G'$ keep their weight. Let $F' \subseteq E(G')$ be an
optimal solution of (POTTS)
for $G'$.

The following Lemma is an easy consequence of the (SUPER) property.

\begin{lem} \label{lem:ex}
There exists an optimal solution of (POTTS) for $G$ which
contains all edges of $F'$.
\end{lem}

{\bf Proof}: Let $F$ be any optimal solution of (POTTS) for $G$.
By the (SUPER) property one has in $G$:

$$f_{G}(F \cup F') \ge f_{G}(F) +  f_{G}(F')-f_{G}(F \cap F').$$

Both $F'$ and $F \cap F'$ are subsets of edges of $G'$ and since $F'$
is optimal in $G'$, $f_{G}(F')= 1+f_{G'}(F') \ge 1+f_{G'}(F \cap F')=
f_{G}(F \cap F')$, whence :

$$f_{G}(F \cup F') \ge f_{G}(F).$$

Since $F$ is optimal in $G$ by hypothesis, we get that $F \cup F'$ is
optimal too, and the Lemma is proved.
\qed

As a consequence of Lemma \ref{lem:ex} we can obtain
an optimal solution $F^{*}$ for $G$
by adding edges to
$F'$. Let
$X_{1}, \ldots, X_{k}$ be the vertices of the connected components
of $G'(F')$, then ${\cal P}_{G}(F')= \{X_{1}, \ldots, X_{k},
\{x\}\}$. Each
connected component in $G(F^{*})$ will be either an element of
${\cal P}_{G}(F')$ or the union of at least two elements of ${\cal
P}_{G}(F')$. The next Lemma shows how the value of a Potts partition
is affected when a subset of connected components are put together.

\begin{lem} \label{lem:tog}
Let ${\cal P}$ and ${\cal P}'$ be two Potts partitions of $G$ such that ${\cal
P}=({\cal P}' \setminus {\cal W}) \cup \{\cup X_{i};X_{i} \in {\cal
W}\}$ for some ${\cal W}\subseteq {\cal P}$. Then
$$f_{G}({\cal P}) = f_{G}({\cal P}') - (|{\cal W}|-1- w(E({\cal W}))),$$
where $E({\cal W})$ denotes the
set of edges of $G$ joining vertices belonging to two different sets
in ${\cal W}$.

In particular if ${\cal P}'$ is optimal then $|{\cal W}|-1- w(E({\cal
W}))\ge 0$ and if ${\cal P}$ is optimal then $|{\cal W}|-1- w(E({\cal
W}))\le 0$.
\end{lem}

{\bf Proof}: Replacing in ${\cal P}$ the sets of ${\cal W}$ by their
union decreases the cardinality of ${\cal P}$ by $|{\cal W}|-1$, on
the other hand let $A$ and $A'$ be the sets of edges induced by the
classes respectively of ${\cal
P}$ and ${\cal P}'$, $A= A' \cup E({\cal W})$ and so the weight of
the edges augment by $w(E({\cal W}))$. The equality is then proved
and the rest follows directly.
\qed

We are now ready to prove the following theorem:
\begin{thm}
\label{thm:W}
For any ${\cal W}\subseteq {\cal P}_{G}(F')$ containing $\{x\}$ and
minimizing $|{\cal W}|-1- w(E({\cal W}))$,
the set $F^{*}=F' \cup E({\cal W})$ is
optimal for (POTTS) in $G$.
\end{thm}

{\bf Proof}:
By Lemma \ref{lem:ex} we know that there exists an optimal solution
$F^{*}$ of
(POTTS) for $G$ which contains $F'$. Let ${\cal W}\subseteq {\cal
P}_{G}(F')$ such that $W=  \cup_{X_{i} \in {\cal
W}} X_{i}$ is an element of ${\cal P}_{G}(F^{*})$. If ${\cal W}$ doesn't
contain $\{x\}$ then, since $F'$ is optimal in $G'$, we get by Lemma
\ref{lem:tog} that $|{\cal W}|-1- w(E({\cal W})) \ge 0$ (this value is
the same in $G$ and $G'$). On the other hand $F^{*}$ is optimal in $G$
which implies that $|{\cal W}|-1- w(E({\cal W})) = 0$, but then $F^{*}
\setminus E({\cal W})$ is also optimal for $G$. Hence there exists an
optimal solution $F^{*}$ of (POTTS) for $G$ containing $F'$ and such that
any element of ${\cal P}_{G}(F^{*})$ not containing $x$ is already in
${\cal P}_{G}(F')$.

So any
${\cal W}\subseteq {\cal P}_{G}(F')$ containing $\{x\}$,  such that
$W=\cup_{X_{i} \in {\cal W}} X_{i}$ induces a connected subgraph of $G$,
and minimizing under these conditions
$|{\cal W}|-1- w(E({\cal W}))$, will provide an
optimal solution $F' \cup E({\cal W})$. (Notice that this minimum is
$\le 0$ since for ${\cal W}= \{\{x\}\}$ we get $0$.)

In order to prove the
theorem it remains only to show that
we may skip the connectivity constraint : we show now that any
subset ${\cal W} \subseteq {\cal P}_{G}(F')$, $\{x\} \in \cal W$, minimizing
$|{\cal W}|-1- w(E({\cal W}))$ corresponds to a connected
subgraph in
$G$. Assume not, and let ${\cal W}_{1}, \ldots, {\cal W}_{l}$ ($l \ge
2$)
be the subsets of ${\cal P}_{G}(F')$ corresponding to the vertices
of the connected components induced by $W=\cup_{X_{i} \in {\cal
W}} X_{i}$ in $G$, ${\cal W}_{1}$
being the one containing $\{x\}$. We have $$|{\cal W}|-1- w(E({\cal
W}))= |{\cal W}_{1}|+ \ldots +|{\cal W}_{l}|-1 - w(E({\cal
W}_{1}))- \ldots -w(E({\cal W}_{l}))$$
For $i \ge 2$, $\{x\} \notin {\cal W}_{i}$ and so as already
noticed $|{\cal W}_{i}|- w(E({\cal
W}_{i})) \ge 1$, which implies  $|{\cal W}|-1- w(E({\cal
W})) > |{\cal W}_{1}|-1 - w(E({\cal
W}_{1}))$ and so we get a contradiction to our assumption on ${\cal
W}$.
\qed

At this point we see that any way to find ${\cal W}$
  will provide a constructive algorithm for getting an optimal solution
of (POTTS) : we start with a solution for a small subgraph (for
example a one vertex subgraph) and then add the vertices one by one,
computing at each step a new extended solution. Fortunately
  an optimal ${\cal W}$ can be computed and in the next section we are
  going to show how.

We remark that the value $|{\cal W}|-1- w(E({\cal W}))$ doesn't depend on the
subgraphs induced by the subsets $X_{i}$ in ${\cal W}$, so we may
ignore these and work in a possibly smaller graph. To be precise :
the result of {\em shrinking} the pairwise disjoint sets
$X_1,\ldots, X_ k$ of ${\cal P}_{G}(F')$ in $G$ is the graph
${\rm shr}(G)=({\rm shr}(V),{\rm shr}(E) )$,
where ${\rm shr}(V)=\{x,x_1,\ldots, x_k\}$
where $x_1,\ldots,x_k$ are distinct new vertices, and the function
${\rm shr}:V\longrightarrow \{x,x_1,\ldots ,x_k\}$ is
defined with ${\rm shr}(v):=x_i$ if $v\in X_ i$ and ${\rm shr}(x):=x$;
the image of an edge $e$ with extremities $x$ and $y$
is ${\rm shr}(e)$ with extremities ${\rm shr}(x){\rm shr}(y)$,
moreover,
if this is a loop (${\rm shr}(x)={\rm shr}(y))$, it is deleted; edges 
keep their
weight, that is $w_{{\rm shr}}({\rm shr}(e))=w(e)$;
sets of vertices or of edges are replaced by the image sets.
There is a one to one correspondence between the subsets ${\cal W}$ of
${\cal P}_{G}(F')$ containing $\{x\}$ and subsets $W$ of vertices of
${\rm shr}(G)$ containing $x$ and for any such $W={\rm shr}({\cal W})$ one has
$|{\cal W}|-1- w(E({\cal W}))=|W|-1- w_{{\rm shr}}(W)$ (as defined in the
preceding chapter $w_{{\rm shr}}(W)$ is the sum of weights of the edges in
the subgraph of ${\rm shr}(G)$ induced by $W$). So ${\cal W}$
will be optimal if and only $W={\rm shr}({\cal W})$ minimizes
$|W|-1- w_{{\rm shr}}(W)$.


\section{A network flow model}\label{sec:PQPW}
Given a graph $H$ with
weight function $w$ on the edges and $W \subseteq V(H)$, let
$b(W)=|W|-1- w(W)$.
 From the previous chapter it is clear that any way of solving the
following problem will provide an algorithm solving (POTTS) :
given a vertex $x$ in $H$,
find $W^*$ a subset of vertices of $H$ containing $x$ such that
$b(W^*)= min (b(W); W \subset V(H), x \in W)$. This
problem is solved both in Picard-Queyranne \cite{PQ} and
Padberg-Wolsey \cite {PW}, using a network flow model that will be
described below. (The work \cite{PQ}, \cite{PW} solve much more
complex problems
involving several network flow computations -- the subroutine
below is the adaptation of an auxiliary procedure.)

We first give some definitions and wellknown facts.
Given a  directed graph and a subset $X$ of its vertices, $\delta
(X)$ will denote the set of arcs leaving $X$, and is called a
{\em cut}; if $s\in X$, $t\notin X$ it is an $(s,t)$-cut;
$\delta(X,Y)$ denotes the set of arcs oriented from $X$ to $Y$.
A function $c$ with nonnegative value on the arcs is called
a {\it capacity} function. If $F$ is a set of
arcs then $c(F):=\sum_{e\in F}c(e)$. A {\it network} is a directed
graph with capacity function. By a well known theorem of Ford and
Fulkerson \cite{FF} a minimum $(s,t)$-cut in a network may be found
by computing
a maximum flow in this network. Very efficient maximum flow
algorithms are known \cite{AMO}.

We now describe in several steps a network $(D, c)=N(H,w)$ that will be
associated to $H$ and $w$:

\noindent 1.
$V(D):=V(H)\cup\{s,t\}$ where $s,t$ are distinct new vertices;

\smallskip\noindent
2. Define for all $u\in V(H)$:  $$p(u):= 1/2\sum_{v\in
V(H),\,uv\in E(H)}  w(uv),  \hbox{ and then }$$
\begin{itemize}
    \item[--]if $p(u)>1$ add an arc $(s,u)$ of capacity $c(s,u)=p(u)-1$,
    \item[--]if $p(u)<1$ then add
    an arc $(u,t)$ of capacity $c(u,t)=1-p(u)$.
\end{itemize}

\smallskip\noindent
3. to each edge $uv$ of $E(H)$ we associate the arcs $(u,v)$
and $(v,u)$ of capacities $c(u,v)=c(v,u)={1\over 2} w(uv)$.

\bigskip
Let $W\subseteq V(H)$, $\delta(\{s\} \cup W)$ is an $(s,t)$-cut of $N(H,w)$ and
reciprocally any $(s,t)$-cut of $N(H,w)$ corresponds to a subset of
vertices of $H$.
There is a close relationship between the capacities of the
$(s,t)$-cuts of $N(H)$ and the values of $b$ on subsets of vertices
of $H$ :
$$c(\delta (\{s\} \cup W))= |W| - w(W) + K=b(W) + K +
1,\eqno{(CUT)}$$
where $K:= c(\delta (s))=\sum_{(s,x)\in A}c(s,x)$.

\bigskip
Indeed, let us see how the capacity of the cuts changes if we
start with the set $\{s\}$ inducing an $(s,t)$-cut of capacity $c(\delta(s))=K$
and then `add' to it
the vertices of $W$ one by one:

The contribution of adding $v$ to the side of $s$ is
$1-p(v)$, for either it decreases
from $p(v)-1$ to $0$ (this happens if $p(v)>1$ see the first case
of `2.' in the construction of $N(H,w)$),
or it increases from $0$ to $1-p(v)$. Thus the
contribution of these arcs is:
    $$\sum_{v\in W} 1- p(v)= |W| - w(W)-
    1/2\sum_{xy \in E(H), x\in W,\,y \notin W}w(xy). $$

On the other hand the contribution of the arcs between $W$ and
$V(H)\setminus W$ is clear: at the beginning it is zero, and at
the end it is:
$$c(W,V(H)\setminus W)=1/2\sum_{xy \in E(H), x\in W,\, y \notin W}w(xy).$$

The change
comparing to $K$ is provided by the sum of the two contributions,
which is $|W| - w(W)$, and so equation (CUT) has been proved.

 From (CUT) we get that any $(s,t)$-cut $C=\{s\} \cup W$ containing the
special vertex $x$ and of minimum capacity
will correspond to our goal : a subset $W$ of vertices of $H$
containing $x$ and
minimizing $b$. Such a cut is easy to obtain : shrink $\{s,x\}$ or
equivalently, add to $N(H,w)$ an arc
of infinite capacity from $s$ to $x$, and compute in the new network a minimum
$(s,t)$-cut : this cut will contain $x$ and will be minimum among
$(s,t)$-cuts containing $x$.

Let us note that this method can be easily {\em generalized to minimize
any modular shift of $-w(W)$}. Adding $|W|-1$ is just a particular
choice of a modular function. That is, one can minimize $(\sum_{w
\in W} m(w))-w(W)$, where $m:V\longrightarrow \Rset$. Indeed, for
generalizing from  $|W|$ to $m(W)$ one only has to write $m(x)$
instead of $1$ in both cases of Step 2 in the construction of N(H,w).

\section{The Algorithm}
\label{sec:alg}
In this section we state the `Optimal Cooperation Algorithm' whose
validity is a consequence of the results of the two preceding chapters.

\medskip\noindent  At each iteration a subset
$U\subseteq V$ and a partition $\cal P$ of $U$ will be at hand. In
Step~0 we give trivial initial values; in Step~1, we choose an
arbitrary vertex $u$ to be added to $U$ and through the following
steps we {\em compute a subset ${\cal W}$ of ${\cal P} \cup \{u\}$
providing a partition of $U \cup \{u\}$ consisting of the
sets contained in $\cal P$ but not in $\cal W$ plus a new class
containing all the other vertices (that is : the new class is the
union of all sets in $\cal W$). The set of edges corresponding to this
new partition is the union of the set of edges corresponding to the
previous one and the edges between the sets in $\cal W$}.

\bigskip
\noindent{\bf OPTIMAL COOPERATION ALGORITHM}

\medskip
\noindent INPUT: A graph $G=(V,E)$, and a weight function $w$:
$E\longrightarrow (0,1)$, where $(0,1):=\{t\in\Rset : 0<t<1\}$.
\smallskip

\noindent OUTPUT: A partition ${\cal P}^*$ optimizing (POTTS) : the
set of edges $A^*$ induced by the sets in ${\cal P}^*$ maximizes
$\{f_{G,w}(A):=c_{G}(A)+ \sum_{e \in A}w(e): A\subseteq E\}$.

\medskip\noindent
0. $U:=\emptyset$, $\Pscr:=\emptyset$.

\medskip\noindent
Do $n$ times  consecutively Steps 1 to 5, and then define
$\Pscr^*=\Pscr$:

\smallskip
\noindent 1. Choose a vertex $u\in V\setminus U$.

\smallskip
\noindent 2. Define $H$ and $w_{H}$ as the result of {\rm shr}inking
the classes of $\cal P$ in $G(U\cup\{u\})$ with weight function
$w$ restricted to the edges of $G(U\cup\{u\})$.

\smallskip
\noindent 3. Construct $N(H,w_{H})$, add an arc of "infinite"
capacity from $s$ to $u$ and using any maximum flow algorithm
compute, in the so obtained network, a minimum $(s,t)$-cut
defined by the set of vertices $C=\{s, u, x_{1}, \ldots, x_{k}\}$ 
where each $x_{i}$
is a vertex of $H$ corresponding to a set $X_{i}\in\cal P$.

\smallskip
\noindent 4. Define $R:=\{ u\} \cup X_1\cup\ldots\cup X_{k}$

\smallskip
\noindent 5. Redefine $U$ and $\cal P$: $U:=U\cup\{u\}$,
$\Pscr := (\Pscr\setminus\{ X_1,\ldots, X_{k}\})\cup \{R\}$;

\smallskip
\noindent END;

\medskip\noindent{\bf Remarks}:

- "Infinite" means big enough to avoid a minimum cut
containing this arc: for example the sum of the capacities of all arcs.

- Note that $C$ can be equal to $\{s,u\}$.

- The graphs $H$ can also be constructed
iteratively with only one shrinking in each iteration, by
adding $u$ and then shrinking the set $C \setminus \{s\}$.

- The choice for $u$ is completely free; this
freedom could be used for making the computations easy.

\begin{thm}
The output $\cal P^*$ is an optimal partition for $(G,w)$.
\end{thm}

\proof
At step 0, $\Pscr:=\emptyset$ is an optimal solution for the subgraph
of $G$ induced by $U:=\emptyset$. Assume now that $\Pscr$ is an optimal
solution for the subgraph
of $G$ induced by $U$, and show that after applying steps 1 to 5 the
new partition is optimal for $U \cup \{u\}$. From the preceding
chapter we know that $W:=C \setminus \{t\}=\{u, x_{1}, \ldots,
x_{k}\}$ is a subset of vertices of $H$ containing $u$  minimizing
$|W|-w_{H}(W)-1$. But $H$ and $w_{H}$ are obtained by shrinking the
classes of $\Pscr$ in $(G(U\cup\{u\}),w)$, hence ${\cal W} =
\{u, X_{1}, \ldots,
X_{k}\}$ is a subset of $\{u\} \cup \Pscr$
minimizing $|{\cal W}|-1- w(E({\cal W}))$  and by Theorem \ref{thm:W}
the partition $(\Pscr\setminus\{ X_1,\ldots, X_{k}\})\cup \{R\}$
is then optimal
\qed

\bigskip
Notice that the "Optimal Cooperation algorithm"  consists merely of $n-1$
network flow computations and $n-1$ shrinking.

\section{Implementation and evaluation}\label{sec:res}
In this section we discuss some aspects of the practical
implementation of the optimal cooperation algorithm.
Here we consider a two dimensional
square lattice where the weights, $w$, follow a bimodal distribution:
\begin{equation}
P(w) =  p \delta(w - w_1) + (1-p) \delta( w-w_2)
\end{equation}
where $\delta(x) =1 $ if $x=0$ and  $\delta(x) =0 $ otherwise.
We restrict ourself to the symmetric case $p=1/2$.
As explained in \cite{FIS} in the infinite lattice limit there is a
second-order phase transition in the model at $w_1+w_2=1$. In the strongly
disordered phase, $w_1+w_2 \ll 1$ the optimal set of edge is $A = \emptyset$,
while in the strongly ordered phase, $w_1+w_2 \gg 1$ it is $A = E$. In
between, the optimal set $A$ has a non-trivial structure and at the transition
point it is a fractal reflecting the critical properties of the
system\cite{FIS}.

As explained in the previous section, for a graph with $n$ vertices
the algorithm consists
in $n-1$ steps. At each step a new site is incorporated to the
current graph for which the optimization has already been performed.
There is a freedom of the order of the added sites, which can be used
to speed up the computation, as well as to obtain the optimal sets
of intermediary lattices. For
example from the computation of the optimal set for an $L \times L$
square lattice with free boundary conditions one can also obtain
the optimal sets of $L-1$ minors of size ranging
from 1 to $L-1$ with free boundary conditions,
which is useful for finite size analysis.

The ``engine'' of the algorithm is the max flow solver,
which is called at each step. We have used the Goldberg
and Tarjan algorithm \cite{RoiDeLaJungle}.
Another important point concerns the building of the network
(point 3 in the Optimal Cooperation Algorithm). It is possible
to build it explicitly as another data structure than the lattice.
Alternatively one can increase up to a sufficiently large value
all the capacities of the edges of a spanning tree of
each connected component as mentioned in the previous
paragraph. In general the cpu time required to
explicitly shrink the connected components is larger than
the cpu time needed to simply increase some capacities. On
the other hand, keeping all vertices will increase the size of the
network in which the max flow has to be found. If all the capacities
are large ($w_1+w_2>1$, that is we are in the ordered phase)
the network will have only two sites at each step, and therefore
the effort to build the network will be largely compensated by
the speed at which the maximum flow is found. In the opposite extreme, when
the optimal set tends to be empty, it is questionable which of
the two strategies is the best. We have found that explicitly
shrinking the connected components is always the most efficient.
As a matter of fact the computation time is about ten times larger
at the critical point $w_1+w_2=1$ than at the ordered phase
($w_1+w_2>1$) and about three times larger than at the disordered
phase ($w_1+w_2<1$).

As an illustration we display in the table below some
typical cpu times needed to find an optimal set on a periodic square
lattice of various size, $L$, at the critical point $w_1+w_2=1$.
 From these results one can extract the empirical rule that in our
algorithm the computation time grows approximately as the fourth power
of the linear size $L$, thus as the second power of the number
of sites. These cpu times refer to a PC with a pentium III,  800 megahertz
processor.

In conclusion in this paper we have presented and implemented a combinatorial
optimization algorithm which solves the optimal cooperation problem in
polynomial time. The procedure, which can be used to obtain the partition
function of the $q$-state random bond Potts model in the large $q$-limit
is, up to our knowledge, the first application of submodular function
optimization in (statistical) physics. Detailed results about the $q$-state
random bond Potts model, such as properties of phase transitions in two-
and three-dimensional lattices will be presented in a separate
publication\cite{future}.

\begin{table}
\begin{center}
\begin{tabular}{|c|c|}
\hline
L&
time\\
\hline
\hline
16 & $\sim$ one second \\
\hline
32& $\sim$ half a minute\\
\hline
64& $\sim$ five minutes\\
\hline
128& $\sim$one hour and a half\\
\hline
256& $\sim$ one  day\\
\hline
512& $\sim$ two weeks\\
\hline
\end{tabular}
\label{matable}
\caption{Typical cpu time (see text)}
\end{center}
\end{table}

\noindent{\bf Acknowledgment}:
The authors are highly indebted
to Maurice Queyranne for answering  a query by e-mail with
pointers to \cite{PQ}, \cite{PW}. Realizing that these provide
only a separation algorithm we made a second query about sharper
algorithms that optimize on (GRAPHIC).  We are thankful to Tom
McCormick for leading us to Ba\"\i ou, Barahona and Mahjoub's paper 
\cite{BBM},
which also opened the doors to other relevant work: Cunningham's
`Attack and Reinforcement' and Barahona's algorithm. Finally we
thank further electronic discussions with Francisco Barahona,
further useful  comments of Maurice Queyranne concerning the
manuscript, and Andr\'as Frank for his courses on submodular
functions.

F.I.'s work has been supported by the Hungarian
National Research Fund under grant No OTKA
TO34183, TO37323, MO28418 and M36803, by the Ministry of 
Education under grant
No. FKFP 87/2001 and by the EC through Centre of Excellence
(No. ICA1-CT-2000-70029).



\begin{thebibliography}{99}

\bibitem{yeomans}
J.M. Yeomans, {\it Statistical Mechanics of Phase Transitions},
Clarendon Press, Oxford (1992).

\bibitem{baxter}
R.J. Baxter,
{\it Exactly solved models in Statistical Mechanics},
Academic Press, New York (1982).

\bibitem{Bspin}
F.~Barahona, R.~Maynard, R.~Rammal, J.~P.~Uhry,
{\it Morphology of ground states of a two-dimensional frustration model},
J.~Phys. {\bf A 15}, 673--679  (1982).

\bibitem{9}
I. Bieche, R. Maynard, R. Rammal and J.P. Uhry,
{\it On the ground states of the frustration model of a spin glass 
by a matching method of graph theory},
J. Phys. {\bf A 13}, 2553-2576, (1980).

\bibitem{MRAdA}
J. C. Angl\`es d'Auriac, R. Maynard,
{\it On the random antiphase state in the \( \pm  \)spin
glass model in two dimensions},
Solid State Communications \textbf{49}, 785 (1984).

\bibitem{GLV}
A.~Galluccio, M.~Loebl, J.~Vondr\'ak,
{\it New Algorithm
for the Ising Problem: Partition  Function for Finite Lattice},
Graphs, Physical Rev. Letters, {\bf 84}, 5924--5927 (2000).

\bibitem{JPhysLett85}
J. C. Angl\`es d'Auriac, M. Preissmann, R. Rammal,
{\it The random field Ising model :
algorithmic complexity and phase transition},
J. Physique Lett. \textbf{46,} 173 (1985).

\bibitem{EuroPhys97}
J. C. Angl\`es d'Auriac and N. Sourlas,
{\it The 3-d Random Field Ising Model at zero temperature},
Europhys. Lett. \textbf{39}, 473 (1997).

\bibitem{JouCompMod97}
J.C. Angl\`es d'Auriac, M. Preissmann and A. Seb\H{o},
{\it Optimal cuts in graphs and statistical mechanics},
Journal of Math. and Computer Modelling \textbf{26,} 1 (1997).

\bibitem{admr}
M.J. Alava, P.M. Duxbury, C.F. Moukarzel and H. Rieger, {Exact
combinatorial algorithms: Ground states of disordered systems}, in
{\it Phase Transitions and Critical Phenomena},
Vol. 18, p 143 (C. Domb and J.L. Lebowitz, eds),
(London, Academic, 2001).

\bibitem{FIS}
R. Juh\'asz, H. Rieger and F. Igl\'oi, 
{\it Random bond Potts model in the large-q limit}, 
Phys. Rev. E {\bf 64}, 56122 (2001).

\bibitem{GLS2}
M. Gr\"otschel, L. Lov\'asz, A. Schrijver,
{\it Geometric Algorithms and Combinatorial Optimization},
Springer Verlag, Berlin Heidelberg (1988).

\bibitem{S}
A. Schrijver,
{\it A Combinatorial Algorithm Minimizing Submodular Functions
in Strongly Polynomial Time},
Journal of Combinatorial Theory, Series B, {\bf 80}, No. 2,
346-355 (2000).

\bibitem{FFI}
S. Iwata, L. Fleischer, and S. Fujishige,
{\it A combinatorial strongly polynomial algorithm for 
minimizing submodular functions},
Journal of the ACM {\bf 48}, Issue 4, 
761-777 (2001).

\bibitem{I}
S. Iwata,
{\it A Fully Combinatorial Algorithm for Submodular Function Minimization}, 
Journal of Combinatorial Theory, Series B {\bf 84}, No. 2, 
203-212 (2002).

\bibitem{FT} A.~Frank, \'E.~Tardos, 
{\it Generalized Polymatroids and Submodular Flows},
Mathematical Programming 42, 489-563 (1988).

\bibitem{CuO}
W.~H.~Cunningham,
{\it Optimal Attack and Reinforcement of a Network},
Journal of the ACM {\bf 32}, No 3, 549-561 (1985).

\bibitem{B}
F.~Barahona,
{\it Separating from the dominant of the spanning tree polytope},
Operations Research Letters {\bf 12}, 201-203 (1992).

\bibitem{BBM}
M.~Ba\"\i ou, F.~Barahona, A.~R.~Mahjoub,
{\it Separation of Partition Inequalities},
Math. of OR, {\bf 25}, 243--254 (2000).

\bibitem{AIPS}
F. Igl\'oi, M. Preissmann, A. Seb\H{o},
{\it Optimal Cooperation},
Cahiers du laboratoire Leibniz (2002).

\bibitem{FWu}
F.Y. Wu,
{\it The Potts model},
Rev. Mod.Phys {\bf 54}, 235 (1982).

\bibitem{KF}
P.W. Kasteleyn and C.M. Fortuin,
J. Phys. Soc. Jpn \textbf{26} (Suppl.), 11 (1969).

\bibitem{Lexercise}
L.~Lov\'asz, {\it Combinatorial Problems and Exercises},
North Holland and Akadémiai Kiad\'o, (1979).

\bibitem{PQ}
J.-C.~Picard, M.~Queyranne,
{\it Selected Applications of Minimum Cuts in Networks},
INFOR 20, 394-422  (1982).

\bibitem{PW}
M.~W.~Padberg, L.~A.~Wolsey,
{\it Trees and Cuts},
Annals of Discrete Mathematics {\bf 17} 511-517 (1983).

\bibitem{FF}
L.R. Ford and D.R. Fulkerson,
{\it Maximum flow through a network}, 
Canad. J. Math {\bf 8}  399--404 (1956).

\bibitem{AMO} R. K. Ahuja, T. L. Magnanti, J. B. Orlin,
Network flows: Theory, Algorithms and Applications, Prenctice-Hall, 1993.

\bibitem{RoiDeLaJungle}
A.V Goldberg and R.E. Tarjan,
{\it A new approach to the maximum-flow problem},
Journal of the Association for Computing Machinery {\bf 35}, 921--940, (1988).

\bibitem{future}
J.C. Angl\`es d'Auriac, F. Igl\'oi, M. Preissmann, A. Seb\H{o}, (unpublished).


\end {thebibliography}
\end {document}